\documentclass[a4paper,11pt]{article}
\usepackage{pos}
\usepackage{graphicx}
\usepackage{booktabs}
\usepackage{amsmath}
\newcommand{\dd}{\mathrm{d}}

\title{Analytic parametrisations of the scale dependence of proton parton distributions}

\ShortTitle{Analytic scale dependence of PDFs}

\author*[a]{David F. Rentería-Estrada}

\affiliation[a]{Instituto de F\'isica Corpuscular, \\ Universitat de Val\`encia--Consejo Superior de Investigaciones Científicas,\\ E-46980 Paterna, Spain}

\emailAdd{david.renteria@ific.uv.es}

\abstract{Parton distribution functions (PDFs) are non-perturbative inputs whose scale dependence is governed by the Dokshitzer--Gribov--Lipatov--Altarelli--Parisi (DGLAP) equations. Their numerical evaluation commonly relies on interpolation of tabulated grids. We discuss an analytic representation of collinear proton PDFs in $(x,Q^2)$, obtained by promoting the parameters of a standard $x$-space parametrisation to functions of the scale. The coefficients are reconstructed from the HERAPDF2.0 NLO set using machine-assisted fitting techniques, with sampling in $10^{-4}\le x\le 0.65$ and scales up to 1~TeV. The resulting ML-PDFs reproduce integrated distributions at the sub-percent level for most flavours, with deviations reaching about 1.5\% for the strange quark at the highest scales. They satisfy valence and momentum sum rules at the percent level and reproduce representative $pp\to\pi$ and $pp\to\gamma\pi$ cross sections within approximately 2\% and 5\%, respectively, in the kinematic regions studied. Compared with grid interpolation through LHAPDF, direct evaluation of the analytic expressions reduces the total Monte Carlo run time by about 11\% and 30\% in these two applications.}

\FullConference{43rd International Conference on High Energy Physics (ICHEP 2026)\\
30 July to 5 August 2026\\
Natal, Brazil\\}

\begin{document}
\maketitle

\section{Introduction}
Predictions for hadron colliders rely on collinear factorisation, in which cross sections are obtained by convolving perturbative partonic cross sections with parton distribution functions. These non-perturbative objects are parametrised at a reference scale $Q_0^2$ using experimental data and evolved to the relevant scale $Q^2$ using the DGLAP equations~\cite{Altarelli:1977zs}. Modern PDF sets, such as HERAPDF2.0~\cite{H1:2015ubc}, are distributed as grids in $(x,Q^2)$ and accessed through interpolation routines provided by LHAPDF~\cite{Buckley:2014ana}. Although accurate, this approach does not yield an explicit analytic expression for the scale dependence, and repeated grid interpolation contributes to the computational cost of Monte Carlo simulations. Following Ref.~\cite{Ochoa-Oregon:2024}, we investigate a direct analytic representation of the $(x,Q^2)$ dependence and test its accuracy and computational performance against the interpolated reference PDFs. Section~2 introduces the parametrisation and fitting procedure. Section~3 examines the resulting distributions, integral errors and sum rules. Section~4 compares hadronic observables and their run times, followed by conclusions in Section~5.

\section{Analytic ansatz and fitting procedure}
Proton PDFs are commonly parametrised at an initial scale $Q_0^2$ using a functional form that captures their small- and large-$x$ behaviour, supplemented by a factor that provides additional flexibility. A representative form is
\begin{equation}
f_i(x,Q_0^2)=N_i\,x^{\alpha_i}(1-x)^{\beta_i}\,P(x,c_{ij}),
\label{eq:basic_pdf}
\end{equation}
where $i$ denotes the parton flavour, $x$ is the fraction of the proton's longitudinal momentum carried by the parton, and $N_i$ is a normalisation factor. The parameters $\alpha_i$ and $\beta_i$ govern the small-$x$ ($x\to 0$) and large-$x$ ($x\to 1$) behaviours, respectively, while $P(x,c_{ij})$ introduces additional flexibility through coefficients $c_{ij}$~\cite{Ochoa-Oregon:2024,H1:2015ubc}. The DGLAP equations determine the evolution from $Q_0^2$ to other scales. Our aim is to construct an explicit analytic representation in $(x,Q^2)$ that can be evaluated without interpolating PDF grids. We assume that the form of Eq.~(\ref{eq:basic_pdf}) remains suitable when its parameters are promoted to functions of $Q^2$~\cite{Ochoa-Oregon:2024},
\begin{equation}
f_i(x,Q^2)=N_i(Q^2)\,x^{\alpha_i(Q^2)}
(1-x)^{\beta_i(Q^2)}P\big(x,c_{ij}(Q^2)\big).
\label{eq:fit_xPDF}
\end{equation}
The resulting analytic approximations, denoted ML-PDFs, reproduce the output of DGLAP evolution embodied in the reference set; they are not new analytic solutions of the evolution equations. We assess both their accuracy and the time required for Monte Carlo calculations using direct ML-PDF evaluations versus LHAPDF interpolation. The coefficients are determined in two steps~\cite{Ochoa-Oregon:2024}. First, the HERAPDF2.0 NLO set~\cite{H1:2015ubc} is sampled at $\mathcal{O}(5000)$ randomly selected points with $x\in[10^{-4},0.65]$ and $Q\in[Q_{\min},1000]$~GeV. The minimum scales are $Q_{\min}=1.37$~GeV for the gluon and light quarks, $1.5$~GeV for charm, and $4.5$~GeV for bottom. The $x$ dependence is fitted at fixed $Q$ to determine the coefficients. Second, the Mathematica routine \texttt{FindFormula} suggests functional forms for their scale dependence, which are used to construct the final analytic parametrisations~\cite{Ochoa-Oregon:2024}. These expressions are available as ancillary files~\cite{ZENODOPDF}. While the valence distributions $u_v$ and $d_v$ are described using Eq.~(\ref{eq:fit_xPDF}), the gluon, sea-quark and heavy-quark distributions require additional flexibility. Inspired by the gluon parametrisation used in HERAPDF2.0, a second term is introduced with a step function. For the gluon, this is written schematically as
\begin{equation}
x\,g(x,Q^2)=f_1(x,Q^2)
-\Theta(x_{C,g}-x)\,f_2(x,Q^2),
\label{eq:gluon}
\end{equation}
where $f_1$ and $f_2$ have scale-dependent coefficients as in Eq.~(\ref{eq:fit_xPDF}), with polynomial factors selected according to the scale region. The Heaviside function activates the second contribution for $x<x_{C,g}$ and removes it for $x>x_{C,g}$. It therefore provides additional freedom in the small-$x$ region without modifying the functional form at larger $x$. The value $x_{C,g}=0.1$ was selected empirically~\cite{Ochoa-Oregon:2024}; it is a fitting choice, not a physical partonic threshold. Similar flavour-dependent cutoffs are used for the sea and heavy quarks. The sharp transition makes it important to examine the approximation near $x_{C,g}$. Because the coefficients also vary substantially with the scale, separate $Q$ regions are fitted: for the gluon, $[Q_0,2.5]$, $[2.5,5]$, $[5,150]$ and $[150,1000]$~GeV~\cite{Ochoa-Oregon:2024}.

To quantify the agreement between ML-PDFs and HERAPDF2.0, we use the integral error
\begin{equation}
\Delta_i(Q^2)=
\frac{I[f_i^{\rm ML},Q^2]-I[f_i^{\rm HERA},Q^2]}
{I[f_i^{\rm HERA},Q^2]},
\qquad
I[f,Q^2]=\int_{10^{-4}}^1\dd x\,f(x,Q^2).
\label{eq:integral_error}
\end{equation}
This signed estimator measures differences between integrated distributions and is less sensitive than pointwise relative errors to fluctuations where the PDF is small. However, positive and negative discrepancies can cancel under integration. Moreover, the integrals extend beyond the sampled fitting interval $x\leq 0.65$. We therefore complement this measure with PDF-shape comparisons, sum-rule tests and predictions for physical observables.

\section{Results}
\begin{figure}[t]
\centering
\includegraphics[width=0.49\linewidth]{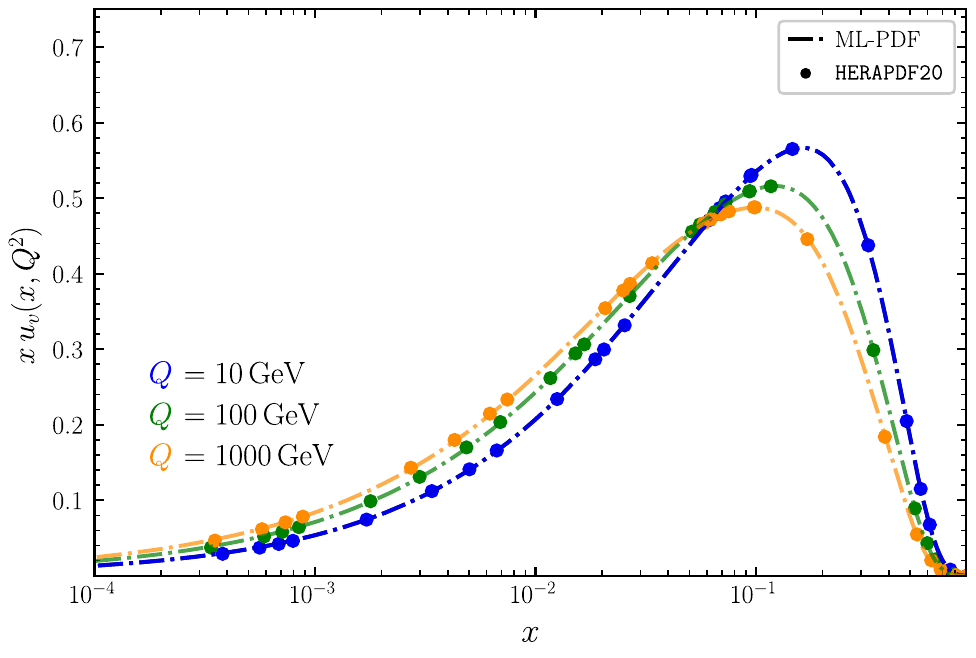}
\includegraphics[width=0.49\linewidth]{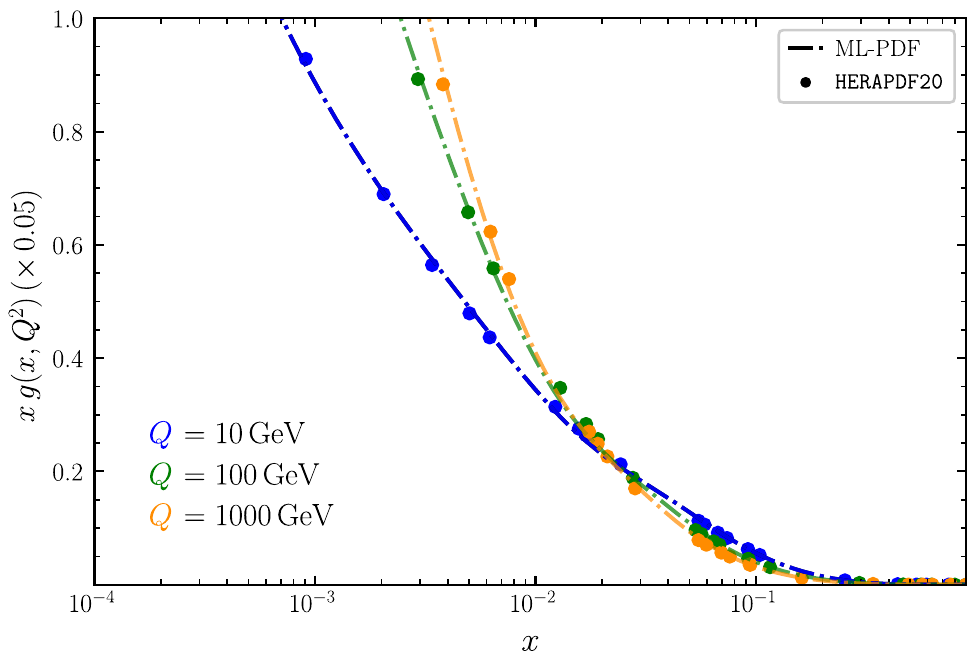}
\caption{ML-PDF approximation (dashed lines) and HERAPDF2.0 (dots) for $x\,u_v$ (left) and $x\,g$ (right, multiplied by 0.05) at $Q=10$, $100$ and $1000$~GeV. }
\label{fig:dist}
\end{figure}

\begin{figure}[t]
\centering
\includegraphics[width=0.49\linewidth]{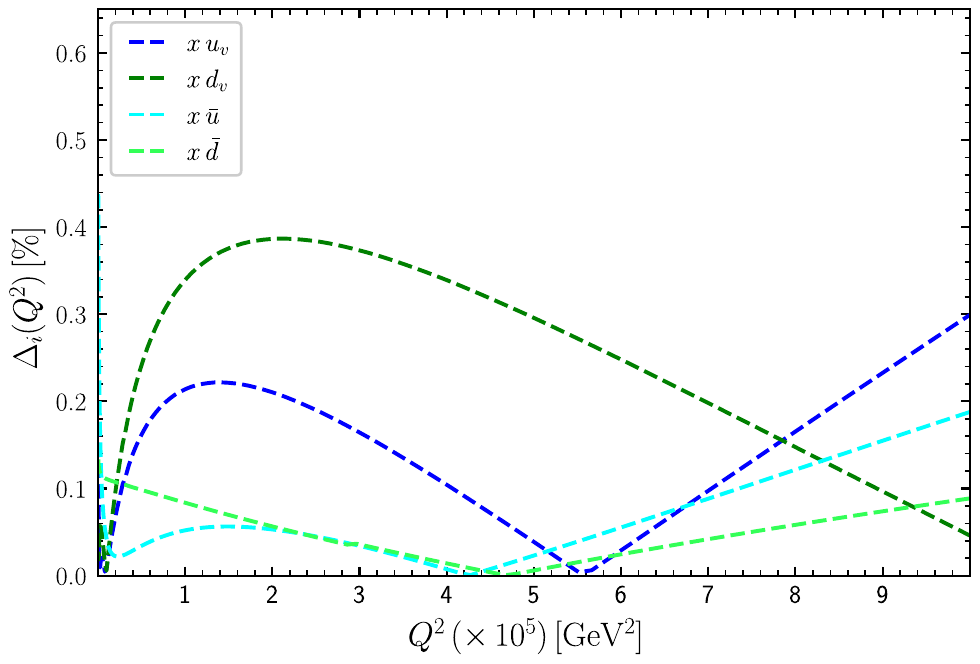}
\includegraphics[width=0.49\linewidth]{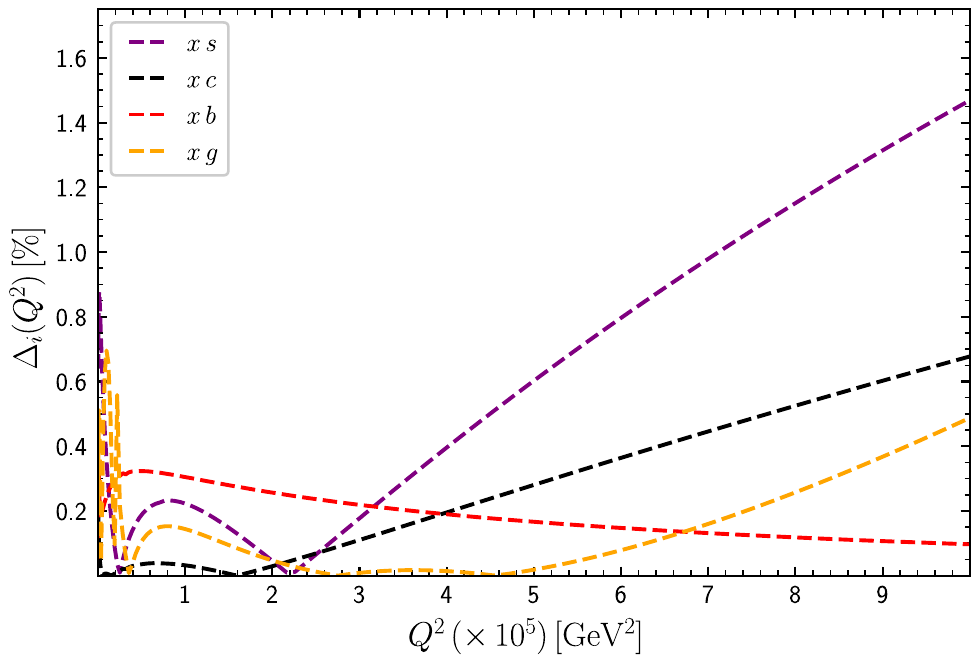}
\caption{Integral error $\Delta_i(Q^2)$ for the dominant light flavours (left) and for $s$, $c$, $b$ and $g$ (right).}
\label{fig:err}
\end{figure}

Figure~\ref{fig:dist} compares the analytic ML-PDF approximations with HERAPDF2.0 for the valence $u_v$ and gluon distributions at $Q=10$, $100$ and $1000$~GeV. Both distributions reproduce the main features of the reference PDFs over the displayed $x$ range, including their variation with the energy scale~\cite{Ochoa-Oregon:2024}.
The valence distribution is well described by a single contribution with a polynomial factor. For the gluon, the additional term in Eq.~(\ref{eq:gluon}) accommodates the more complex small-$x$ behaviour. Some deviations remain visible there, although the main features of the shape and scale dependence are reproduced. This shape comparison is important because a small integrated error need not imply uniform pointwise agreement. Figure~\ref{fig:err} displays the integral error of Eq.~(\ref{eq:integral_error}) for each flavour. Most distributions have deviations below $0.5\%$, while charm reaches roughly $0.7\%$ and the strange-quark distribution exhibits the largest deviation, approximately $1.5\%$ at $Q\simeq 1000$~GeV~\cite{Ochoa-Oregon:2024}.

These results demonstrate sub-percent accuracy for most integrated distributions. They do not, however, directly bound the error of an arbitrary cross section: partonic kernels weight different $x$ regions and need not preserve the cancellations in Eq.~(\ref{eq:integral_error}). The valence and momentum sum rules provide an additional consistency test. The conditions $\int_0^1\dd x\,u_v=2$ and $\int_0^1\dd x\,d_v=1$ are reproduced at the $0.1$--$0.2\%$ level. The momentum sum rule is satisfied within about $0.6\%$ for $Q<60$~GeV, with deviations reaching approximately $1.5\%$ near $Q=500$~GeV~\cite{Ochoa-Oregon:2024}. The latter behaviour is driven mainly by the gluon contribution. These sum rules were not explicitly imposed in the fits, so their approximate fulfilment is a useful cross-check rather than a guarantee of exact conservation.

\section{Hadronic observables and computational performance}
To assess the phenomenological impact of the analytic approximations, we consider two hadronic observables calculated using the DEHSS2014 fragmentation functions~\cite{PhysRevD.91.014035}: single-inclusive pion production, $pp\to\pi$, at $\sqrt{S}=7$~TeV, and associated photon--pion production, $pp\to\gamma\pi$, at $\sqrt{S}=13$~TeV. The Monte Carlo integrations employ $10^5$ and $10^8$ points, respectively. The latter process is evaluated at NLO accuracy in QCD~\cite{deFlorian:2010vy,Renteria-Estrada:2021rqp}.
\begin{figure}[t]
\centering
\includegraphics[width=0.45\linewidth]{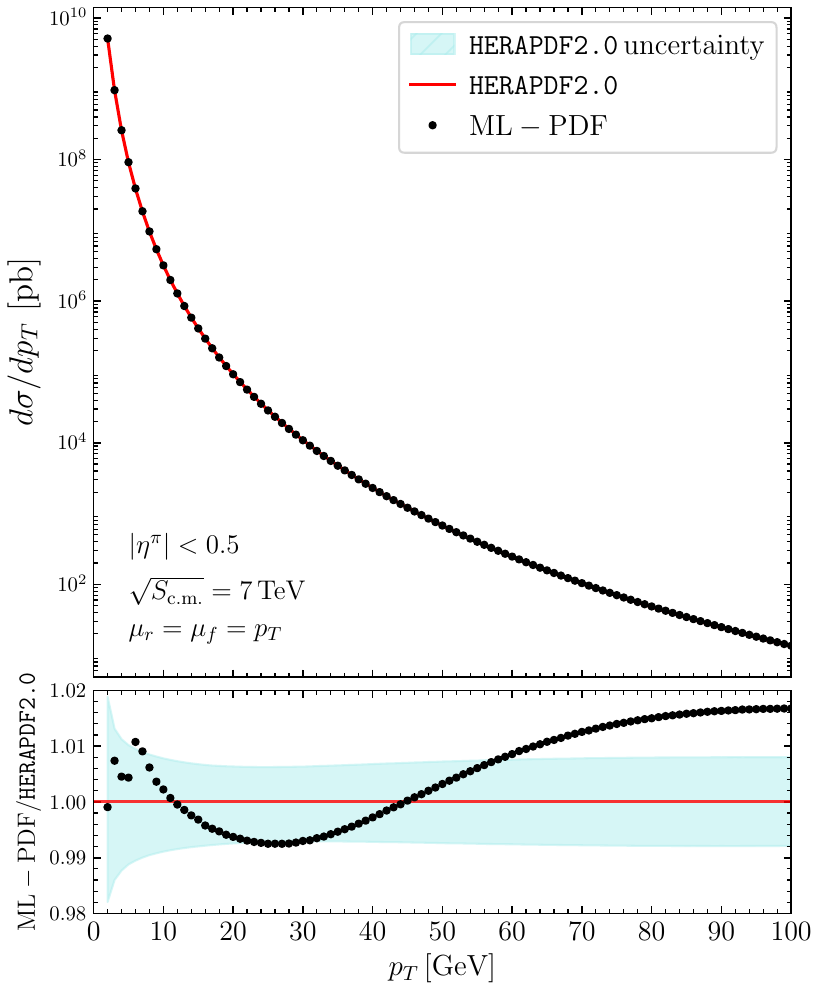}
\hspace{0.03\linewidth}
\includegraphics[width=0.45\linewidth]{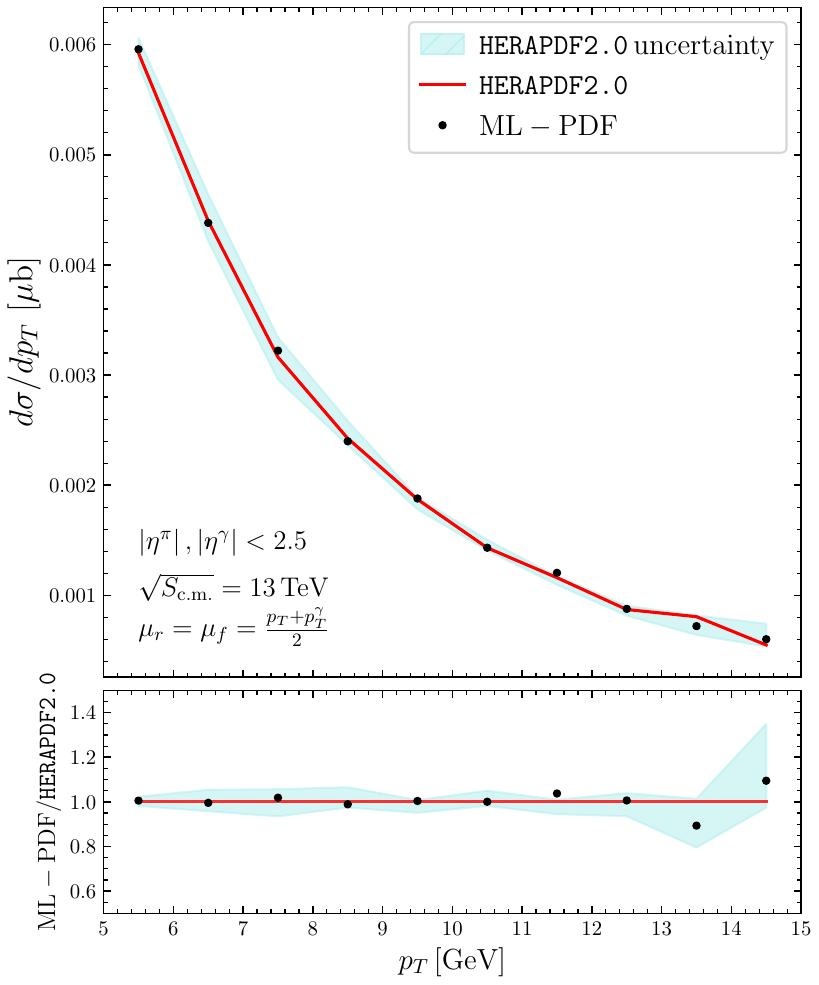}
\caption{Cross sections for $pp\to\pi$ at 7~TeV (left) and $pp\to\gamma\pi$ at 13~TeV (right), computed with HERAPDF2.0 and ML-PDFs. The lower panels show ratios to the HERAPDF2.0 predictions, while the light-blue bands represent propagated PDF uncertainties.}
\label{fig:obs}
\end{figure}
The predictions obtained with ML-PDFs are compared with those using the original HERAPDF2.0 grids in Fig.~\ref{fig:obs}. For single-inclusive pion production, the differences remain well below $2\%$ and are mostly contained within the HERAPDF2.0 uncertainty band. For photon--pion production, the deviations are within $5\%$ for $p_T\leq 10$~GeV, where the cross section is largest. The corresponding predictions are compatible with the propagated PDF uncertainties~\cite{Ochoa-Oregon:2024}. These comparisons provide a more direct phenomenological validation than the integral error alone. Although the ML-PDFs do not reproduce every point of the reference distributions exactly, the deviations in these observables remain controlled. This supports their potential use in numerical calculations for which repeated evaluations of PDFs represent a non-negligible computational cost. An additional advantage of the analytic representation is that PDF evaluations no longer require loading and interpolating tabulated grids. The execution times obtained for both processes are summarised in Table~\ref{tab:speed}. Relative to LHAPDF, the analytic ML-PDFs reduce the total run time by approximately $11\%$ for $pp\to\pi$ and $30\%$ for $pp\to\gamma\pi$~\cite{Ochoa-Oregon:2024}.

\begin{table}[t]
\centering
\begin{tabular}{lccc}
\toprule
Observable & LHAPDF (s) & ML-PDFs (s) & Gain (\%)\\
\midrule
$pp\to\pi$ & 628.320 & 558.854 & 11.06\\
$pp\to\gamma\pi$ & 12452.273 & 8671.827 & 30.36\\
\bottomrule
\end{tabular}
\caption{Execution times for the two hadronic observables using LHAPDF and the analytic ML-PDFs~\cite{Ochoa-Oregon:2024}.}
\label{tab:speed}
\end{table}

The gains depend on the numerical structure of each calculation and the frequency of PDF evaluations. Both benchmarks show that direct analytic evaluation can reduce the total integration time while remaining consistent with the reference predictions at the level of the HERAPDF2.0 uncertainties in the regions studied.

\section{Conclusions}
We have presented an explicit analytic approximation to the $(x,Q^2)$ dependence of collinear proton PDFs, obtained by making the parameters of an $x$-space ansatz scale dependent and fitting them to HERAPDF2.0~\cite{Ochoa-Oregon:2024}. Most integrated distributions agree with the reference set at the sub-percent level, while the strange-quark deviation reaches approximately $1.5\%$ at $Q \approx 1000$ GeV. The valence and momentum sum rules hold at the percent level, and the two hadronic benchmarks agree with the reference predictions within about $2\%$ and $5\%$ in the respective regions considered. Direct ML-PDF evaluation also reduces the total Monte Carlo run time by $11\%$ and $30\%$ in these benchmarks. These results establish the approach as a useful proof of concept, rather than a replacement for PDF extraction or a new solution of DGLAP evolution. The same strategy could be explored for other PDF sets and fragmentation functions; recent variational quantum-circuit representations of pion fragmentation functions provide another possible direction for compact non-perturbative parametrisations~\cite{renteriaFF}.

\paragraph{Acknowledgments.} \sloppy This work is supported by the Spanish Government and ERDF/EU - Agencia Estatal de Investigaci\'on (MCIN/AEI/10.13039/\allowbreak 501100011033) Grants No.\ PID2023-146220NB-I00 and CEX2023-001292-S, by Generalitat Valenciana Grants No.\ CISPINCOM/2026/004 and CIGRIS/2022/145, by the Quantum Spain project and by the CSIC Interdisciplinary Thematic Platform on Quantum Technologies (PTI-QTEP+).

\end{document}